\documentclass[aps,prd,preprintnumbers,showpacs,superscriptaddress,nofootinbib,amsmath,amssymb,floats,floatfix,showkeys,notitlepage,longbibliography]{revtex4-1}
\usepackage{comment}
\usepackage{graphicx}
\usepackage{subfigure}
\usepackage{palatino}
\usepackage[commandnameprefix=always]{changes}
\usepackage{hyperref}
\hypersetup{colorlinks=true,linkcolor=red,urlcolor=red,citecolor=red}
\usepackage[toc,page]{appendix}
\usepackage[normalem]{ulem}

\usepackage{lipsum}
\usepackage{graphicx}
\usepackage{subfigure}
\usepackage{palatino}
\usepackage{float}
\usepackage{sans}
\usepackage{adjustbox}
\usepackage{latexsym}
\usepackage{amsmath}
\usepackage{amssymb}
\usepackage{amsfonts}
\usepackage{dcolumn}
\usepackage{bm}
\usepackage{tikz}
\usepackage{bigints}
\usepackage{array,tabularx,multirow,booktabs}
\usepackage[tracking=true]{microtype}
\SetTracking{}{500}
\SetTracking{encoding={*}, shape=sc}{40}
\allowdisplaybreaks
\usepackage{adjustbox}
\usepackage{latexsym}
\usepackage{amsmath}
\usepackage{amssymb}
\usepackage{amsfonts}
\usepackage{dcolumn}
\usepackage{bm}
\usepackage{tikz}
\usepackage{bigints}
\usepackage{array,tabularx,multirow,booktabs}
\usepackage[tracking=true]{microtype}
\usepackage{color}
\allowdisplaybreaks

\begin{document}

\title{Cosmic-Quantum Connections: Assessing the Viability of Weak Gravity and Weak Cosmic Censorship Conjectures in Kerr-Newman-Kiselev-Letelier Black Hole}

\author{Saeed Noori Gashti}
\email{saeed.noorigashti@stu.umz.ac.ir; saeed.noorigashti70@gmail.com}
\affiliation{School of Physics, Damghan University, P. O. Box 3671641167, Damghan, Iran}

\author{Behnam Pourhassan}
\email{b.pourhassan@du.ac.ir, b.pourhassan@candqrc.ca}
\affiliation{School of Physics, Damghan University, P. O. Box 3671641167, Damghan, Iran}
\affiliation{Center for Theoretical Physics, Khazar University,
41 Mehseti Street, Baku, AZ1096, Azerbaijan}

\vspace{1.5cm}\begin{abstract}
This paper addresses a potential validation of the weak gravity conjecture (WGC) with the weak cosmic censorship conjecture (WCCC), as a significant challenge in quantum gravity. We explore the viability of the WGC and WCCC in the context of the Kerr-Newman-Kiselev-Letelier (KNKL) black hole. Although these conjectures appear unrelated, but surprising connection between these conjectures, It establishes a bridge between the quantum and the cosmic. By imposing specific constraints on the black hole's parameters, we demonstrate that the WGC and WCCC can be compatible in certain regions. We examine the properties of the KNKL black hole for $q/m > (Q/M )_{ext}$, where $(Q/M )_{ext}$ is the charge-to-mass ratio of a large extremal black hole. We present figures to test the validity of both conjectures simultaneously. Without the spin parameter \(a\), the cloud of string parameter \(b\), quintessence parameter \(\gamma\), and equation of state parameter \(\omega\), the black hole either has two event horizons if \(Q^2/M^2 \leq 1\) or none event horizon if \(Q^2/M^2 > 1\) which leads to a naked singularity that contradicts the WCCC. However, when \(a\), \(b\), \(\gamma\), and \(\omega\) are present, the black hole has event horizons in some regions in the \(Q^2/M^2 > 1\) that ensure the singularity is covered and both the WGC and WCCC are fulfilled. Actually, we face this issue in the extremality state of the black hole viz these conjectures remain viable, with the black hole maintaining an event horizon. We conclude that certain regions of \(a\), \(b\), \(\gamma\), and \(\omega\) parameters can make the WGC and WCCC compatible, indicating their agreement when these parameters are present.
\end{abstract}

\date{\today}



\maketitle
\tableofcontents
\newpage
\section{Introduction}
Quantum gravity is a captivating and complex field that has garnered significant interest from researchers across various disciplines. One of the prominent research efforts in this domain is the swampland program, which seeks to uncover the universal principles that any consistent theory of quantum gravity must adhere to. The central premise of the swampland program is that not all low-energy effective field theories can be embedded within a quantum theory of gravity, such as string theory. This program aims to delineate the criteria that separate theories that can be part of the string landscape from those that belong to the swampland \cite{1,2,3,4,5,6}. The motivation for the swampland program is multifaceted, drawing from diverse areas such as black hole physics, the AdS/CFT correspondence, and string theory constructions. By exploring these areas, the swampland program aspires to illuminate the fundamental nature of quantum gravity and its far-reaching implications for cosmology and particle physics \cite{1,2,3,4,5,6}. A pivotal criterion in the swampland program is the absence of global symmetries in quantum gravity, while gauge symmetries are permitted. This criterion leads to the  WGC, which posits that there must exist particles with a charge-to-mass ratio greater than one ($q/m > 1$) in any quantum theory of gravity. This implies that gravity is the weakest force in all interactions. The WGC is one of several conjectures within the swampland program that help identify effective field theories consistent with quantum gravity \cite{1,2,3,4,5,6,7,8}. For further details on the swampland program and its various conjectures, we are encouraged to explore the extensive literature on the subject. Additionally, the swampland program's relevance extends to various cosmological concepts, including black hole physics, thermodynamics, black brane solutions, and cosmological inflation, offering a rich tapestry of research avenues for those interested in the profound questions surrounding quantum gravity \cite{a,b,c,d,e,f,g,h,i,j,k,l,m,n,o,p,q,r,s,t,u,v,w,x,y,z,aa,bb,cc,dd,ee,ff,gg,hh,ii,jj,kk,ll,mm,nn,oo,pp,qq,rr,ss,tt,uu,vv,ww,xx,yy,zz,aaa,bbb,ccc,ddd,eee,fff,ggg,hhh,iii,jjj,kkk,lll,mmm,nnn,ooo,ppp,qqq,rrr,sss,ttt,uuu,vvv,www,xxx,yyy,zzz}. \\\\ These concepts are derived from past works in string theory, black hole thermodynamics, and the study of dualities like AdS/CFT. Historical research in these areas has laid the groundwork for the swampland program, providing the theoretical tools and frameworks necessary to explore the fundamental nature of quantum gravity. By building on these past achievements, the swampland program aims to address some of the most profound questions in modern physics.
Another pivotal concept in theoretical physics is the WCCC, proposed by Roger Penrose to address the paradoxes arising from the existence of singularities in general relativity \cite{9}. The WCCC posits that singularities resulting from gravitational collapse are always concealed behind event horizons, thereby preserving the causal structure and predictability of the theory.
However, a notable challenge arises when considering the WGC alongside the WCCC, particularly in the context of the Reissner–Nordström (RN) black hole. The RN black hole, a solution to the Einstein–Maxwell equations, describes a charged black hole. When the charge \(Q\) exceeds the mass \(M\) of the black hole (\(Q > M\)), the RN solution violates the WCCC. Conversely, when the RN black hole decays to an extremal state where \(Q = M\), energy conservation implies the existence of decay products with a charge-to-mass ratio greater than one. In this scenario, the WGC is satisfied, but the WCCC is violated. This presents a significant challenge for the WGC \cite{10}.\\

Recent studies have explored the WCCC in greater depth, revealing that the presence of dark matter and a cosmological constant can prevent the overcharging of black holes, a scenario not possible in a vacuum. Overcharging can only occur if there is an exact equilibrium between the influences of dark matter and the cosmological constant.
In this paper, we address the challenge of reconciling the WGC and the WCCC by examining the KNKL black hole. We investigate the properties of these black holes in their extremal state and demonstrate that by imposing certain constraints on the parameters of the metric, such as \(a\), \(b\), \(\gamma\), and \(\omega\), we can achieve compatibility between the Mild WGC and the WCCC. Additionally, we explore the case where \(Q > M\) and present intriguing points to test the simultaneous validity of the Mild WGC and the WCCC \cite{10000}. Our research is motivated by recent observational evidence of the accelerated expansion of the universe \cite{11,12}, which suggests the existence of a repulsive gravitational force due to negative pressure on cosmological scales. One possible explanation for this phenomenon is the presence of a cosmological constant, corresponding to the vacuum energy in Einstein's equations. However, the cosmological constant also influences the properties of black holes and their compatibility with the Mild WGC and the WCCC. Therefore, we examine how these black holes can account for the consistency of quantum gravity, considering the Mild WGC and its effect on the validation of both the Mild WGC and the WCCC. The primary motivation of this article is to explore the possibility of resolving the theoretical conflict between the Mild WGC and the WCCC, which has significant implications for quantum gravity and black hole physics. This challenging and intriguing research topic could illuminate some fundamental aspects of nature, offering new insights into the interplay between these conjectures and the broader framework of theoretical physics.\\
The structure of this paper is as follows:
Section II: We provide a comprehensive review of the fundamental characteristics of the KNKL black hole, along with the implications of the Mild  WGC and the WCCC. Also, We investigate the compatibility between the Mild WGC and the WCCC within the framework of the KNKL black hole. Also, we examine the properties and constraints of the relevant parameters, presenting illustrative figures to evaluate the simultaneous validity of the Mild WGC and the WCCC.
Section III: We summarize our findings and conclude with final remarks, highlighting the key insights and potential future directions for research.
\section{The Model}
\begin{figure}[H]
 \begin{center}
 \subfigure[]{
 \includegraphics[height=5cm,width=4cm]{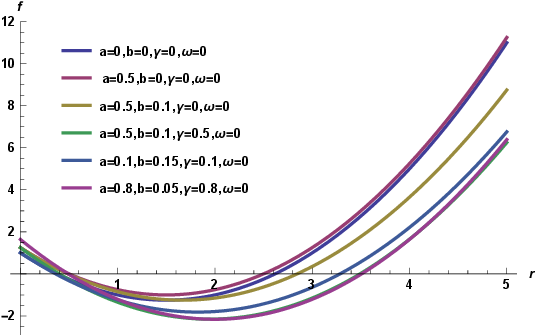}
 \label{1a}}
 \subfigure[]{
 \includegraphics[height=5cm,width=4cm]{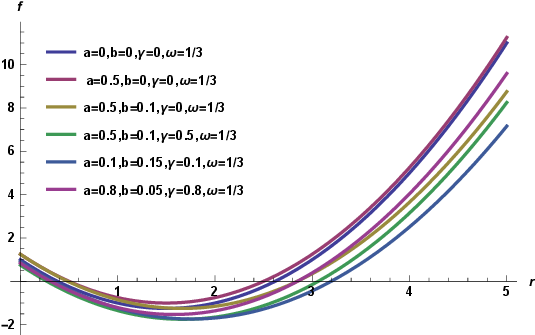}
 \label{1b}}
 \subfigure[]{
 \includegraphics[height=5cm,width=4cm]{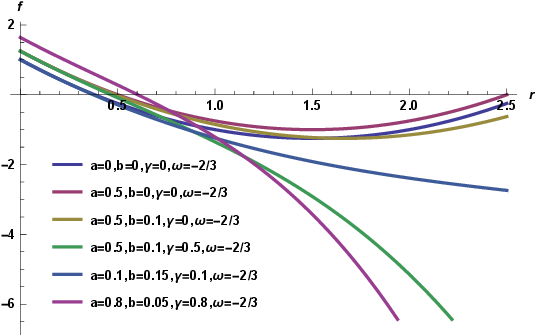}
 \label{1c}}
 \subfigure[]{
 \includegraphics[height=5cm,width=4cm]{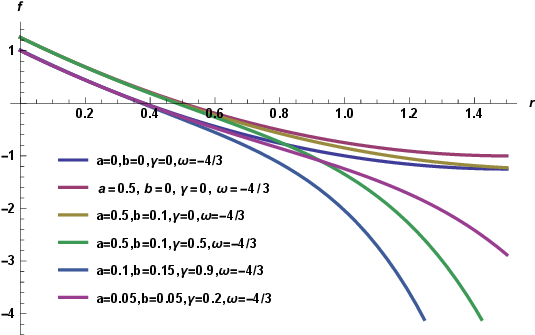}
 \label{1d}}
  \caption{\small{The metric function of the KNKL black hole}}
 \label{m1}
 \end{center}
 \end{figure}

An extremal black hole necessitates the existence of superextremal particles with a charge-to-mass ratio greater than one. If such a black hole can exist without creating observable naked singularities, it indicates a stable balance of forces that constitutes black hole mechanics.
Various conditions such as an event horizon, the WCCC, stability, thermodynamic properties, and more can be examined to analyze a black hole with general behavior. To find more evidence for the WGC, we must study various cosmic phenomena and contradictions, and compare them with existing observable data. This approach leads to a special classification of black holes from the perspective of the WGC. Another important aspect is understanding and applying thermodynamic relationships and respect for the laws of thermodynamics. This work creates a beautiful bridge between quantum mechanics and cosmology by investigating the WGC and the concept of cosmic censorship.
To gather more evidence for the WGC, researchers need to examine a wide range of cosmic cases. This involves identifying and analyzing contradictions within observable data. Classifying these cases helps them understand how different types of black holes behave under the conjecture's framework \cite{1}.
Understanding the thermodynamic properties of black holes is crucial. This includes memorizing and applying the laws of thermodynamics to black hole systems. These relationships provide insights into the stability and decay processes of black holes, which are essential for testing the WGC. The pursuit of evidence for the WGC involves a comprehensive study of cosmic phenomena, thermodynamic properties, and the interplay between quantum mechanics and cosmology. This integrated approach not only tests the validity of the WGC but also enhances our understanding of the universe's fundamental principles. We should note that in normal and sub-extremal modes, We can't consider the WGC criteria. However, the structure and examination of the WGC begin with the assumption of extremal and super-extremal states. This leads to the conclusion that the simultaneous establishment of the two conjectures in most black holes is not possible, necessitating a special classification for black holes.
In general, black holes that are somehow included in the structure of string theory, such as Gauss-Bonnet black holes, Braneworld black holes, and charged black holes with structures such as quintessence, string cloud, perfect fluid, and others, can be categorized in this way. This seems to be a very interesting task, providing a new classification of black holes to examine other evidence and prove the WGC \cite{1}.
Charged black holes (with quintessence, perfect fluid, string cloud, etc) in extremal or superextremal states in our universe can decay by emitting electrons, which are superextremal particles with a charge-to-mass ratio greater than one. This emission process satisfies the  WGC because it ensures that gravity remains the weakest force. Therefore, the choice of the KNKL black hole appears to be highly appropriate. The roles of the quintessence parameter, which represents the influence of dark energy, and the cloud of strings in this black hole are significant. These elements can satisfy the WGC by influencing the structure of the black hole while maintaining the conditions of the WCCC. If a suitable range of compatibility between the WGC and WCCC can be established, this black hole could be classified into a new category based on the . This would open avenues for further investigation and potentially provide more evidence supporting these theoretical frameworks. The KNKL black hole is a solution to the Einstein field equations that generalize the Kerr–Newman black hole by incorporating additional parameters to account for quintessence and a cloud of strings. Here are some key aspects of the KNKL black hole. The KNKL black hole metric extends the Kerr–Newman metric by including terms that represent the effects of quintessence and a cloud of strings. The metric is given by \cite{13,14},
\begin{equation}\label{eq1}
ds^2 = -\frac{F(dt - a \sin^2 \theta d\phi)^2}{\Sigma} + \frac{\Sigma}{F} dr^2 + \Sigma d\theta^2 + \sin^2 \theta \frac{(adt - (r^2 + a^2)d\phi)^2}{\Sigma}.
\end{equation}
In this metric, the coefficients are defined as,
\begin{equation}\label{eq2}
\Sigma = r^2 + a^2 \cos^2 \theta,
\end{equation}
and
\begin{equation}\label{eq3}
F = (1 - b)r^2 + a^2 + Q^2 - 2Mr - \gamma r^{-3\omega + 1}.
\end{equation}
Where we face with \(a\) spin parameter, \(Q\) and \(M\) charge and mass parameter of the black hole, \(\gamma\) quintessence parameter, which represents the influence of dark energy, \(b\) cloud of strings (CS) parameter, which accounts for the presence of a string cloud around the black hole and \(\omega\) equation of state parameter. Note that M is the mass parameter and the ADM mass of the system can be obtained by rescaling the coordinates \cite{15}. The horizons of the KNKL black hole are determined by solving \(F = 0\). The presence of the quintessence parameter \(\gamma\) and the CS parameter \(b\) affects the size and structure of the horizons. Generally, increasing \(\gamma\) and \(b\) leads to a larger horizon radius compared to the Kerr black hole. The KNKL black hole provides a more comprehensive model for studying the effects of dark energy and string clouds on black hole properties. It helps in understanding how these additional factors influence the geometry and physical characteristics of black holes \cite{13,14}.\\
We examine the influence of the  \(\gamma\), \(b\),  \(a\), $\omega$, $Q$, and mass $M$ on the viability of WGC-WCCC for the mentioned black hole. The horizon of the black hole spacetime is found by setting \(F = 0\).  For a more detailed discussion on the horizon structure of the KNKL black hole, see \cite{14}.
The metric function of the KNKL black hole associated with Eq. (\ref{eq3}) is well illustrated in Fig. (\ref{m1}) for different values of the free parameters, including the case where \( M > Q \).
\subsection{ WGC-WCCC Viability}
We first consider the metric of the Kerr–Newman black hole, which is a solution to the Einstein-Maxwell equations representing a rotating, charged black hole. To determine the locations of the event horizons for such a black hole, we solve the equation \( f(r) = 0 \).
In the context of the Kerr–Newman black hole, if the charge \( Q \) exceeds the mass \( M \) (\( Q > M \)), the black hole does not possess an event horizon. Consequently, the singularity becomes exposed to external observers, a phenomenon known as a naked singularity. This situation contravenes the WCCC, which posits that singularities should be concealed behind event horizons to prevent them from being observed. To address this inconsistency, we have selected the KNKL black hole. By applying the relevant equations, specifically Eq. (\ref{eq3}), we can analyze the properties and behavior of this black hole.
\begin{equation}\label{eq4}
a^2+(1-b) r^2-2 M r+Q^2=\gamma r^{1-3 \omega}
\end{equation}
Eq. (\ref{eq4}) presents a significant challenge to solve. To address this, we employ various plots and analyze different scenarios to find their solutions. We begin with the case of asymptotically flat space and transition it to an extremality state. Based on Eq. (\ref{eq4}), we sketch two curves, \( F_L \) and \( F_R \), and label them as Figs. (\ref{m3}) to (\ref{m6}). We consider two scenarios: \( Q^2/M^2 \leq 1 \) and \( Q^2/M^2 > 1 \). In Fig. (\ref{3a}), (\ref{4a}), (\ref{5a}) and (\ref{6a}), when \( F_R = 0 \), the curve \( F_L \) intersects the \( r \)-axis at two points, \( (r_-, r_+) \). These points represent the inner and outer event horizons of the standard KNKL black hole. To find the values of \( r_0 \), \( b_{ext} \) and \( \gamma_{ext} \), we use Eq. (\ref{eq4}) along with the condition that the curves \( F_R \) and \( F_L \) are tangent to each other at the point \( r_0 \). This tangency implies that the slopes of the two curves are equal at \( r_0 \). By solving for \( b_{ext} \) and \( \gamma_{ext} \) under this condition, we can determine the values of \( b_{ext} \) and \( \gamma_{ext} \) and \( r_0 \) using the following expressions:
\begin{equation}\label{eq5}
a^2+(1-b) r_0^2-2 M r_0+Q^2=\gamma r_0^{1-3 \omega}
\end{equation}
and
\begin{equation}\label{eq6}
2 (1-b) r_0-2 M=\gamma (1-3 \omega) r_0^{-3 \omega}
\end{equation}
So, with respect to the Eqs. (\ref{eq5}) and (\ref{eq6}), we will have,
\begin{equation}\label{eq7}
r_0=\frac{1\pm \sqrt{1-4 \left(-a^2-Q^2\right) (-3 b \omega-b+3 \omega+1)}}{2 (3 b \omega+b-3 \omega-1)}
\end{equation}
Additionally, we determine \( b_{ext} \) and \( \gamma_{ext} \) by applying a series of simplifications,
\begin{equation}\label{eq8}
b_{ext}= r_0^{-3 \omega-2} \left(a^2 r_0^{3 \omega}-\gamma r_0-2 m r_0^{3 \omega+1}+Q^2 r_0^{3 \omega}+r_0^{3 \omega+2}\right)
\end{equation}
\begin{equation}\label{eq80}
\gamma_{ext}=-r_0^{3 \omega-1} \left(-a^2+b r_0^2+2 M r_0-Q^2-r_0^2\right)
\end{equation}
To achieve the extremal state of a black hole, we maintain constant parameters and increase \( Q^2/M^2 \) until it reaches \( (Q^2/M^2)_{ext} \). This condition ensures the black hole has a single event horizon. Alternatively, we can approximate the extremal black hole by fixing the free parameters and applying Eqs. (\ref{eq5}), (\ref{eq6}), (\ref{eq7}), (\ref{eq8}), (\ref{eq80}) and mass this yields the relation for the extremal black hole as follows,
\begin{equation}\label{eq9}
\begin{split}
&(\frac{M^2}{Q^2})_{ext}=\gamma^{-(3 \omega+1)} \left(\frac{1+ \sqrt{1-4 (b-1) (3 \omega+1) \left(1+a^2/Q^2\right)}}{(b-1) (3 \omega+1)}\right)^{-3 \omega}\\&-\frac{(b-1) (3 \omega+1) \left(1+a^2/Q^2\right)}{1+ \sqrt{1-4 (b-1) (3 \omega+1) \left(1+a^2/Q^2\right)}}
\end{split}
\end{equation}
By applying some simplifications and performing a series of straightforward calculations, we can rewrite the relation for \( \left(\frac{Q^2}{M^2}\right)_{ext} \) in the following form,
\begin{equation}\label{eq10}
(\frac{Q^2}{M^2})_{ext}=1+\bigg[\gamma^{2/{3 \omega+1}} (b-1)^{-3 \omega}\div\frac{1}{2} a^2 (b-1)(3 \omega+1)\bigg]
\end{equation}
Thus, with respect to some regions of free parameters (\(a\), \(b\), \(\gamma\), and \(\omega\)), we obtain as,
\begin{equation}\label{eq100}
(\frac{Q^2}{M^2})_{ext}=1+\delta, 
\end{equation} 
where $\delta$ is a constant positive value. Now, we want to investigate whether the WGC and the WCCC are satisfied by the KNKL black hole in the extremal state. The simultaneous examination of the WGC and the WCCC is intriguing because both conjectures address fundamental aspects of gravity and black holes, and their interplay can provide deeper insights into the nature of spacetime and singularities.
Examining these conjectures together can help test the consistency of theories of quantum gravity and explore whether the presence of charged particles (as required by the WGC) can prevent the formation of naked singularities, thereby supporting the WCCC.\\
Not all black holes exhibit properties that directly test these conjectures simultaneously. The focus is primarily on extremal and near-extremal charged black holes (With PFDM, quintessence, and the cloud of strings, etc) because they are more likely to meet these conjectures. Also, The WGC and the WCCC are two distinct ideas in theoretical physics, each addressing different aspects of fundamental physics. While both conjectures involve the term ”weak” and pertain to fundamental aspects of gravity and spacetime, they are distinct in their focus and implications. The WGC primarily concerns the relative strengths of forces in a quantum gravity framework, while the WCCC deals with the behavior of singularities and event horizons in gravitational collapse scenarios within classical general relativity. Combining these concepts, one might speculate about potential connections between the WGC and cosmic censorship, such as whether the WGC could have implications for cosmic censorship scenarios. The connection between these two conjectures and their common points related to different physical concepts can serve as a bridge for better communication between general relativity and quantum mechanics, or quantum gravity. This connection clearly provides intuition and documentation for further investigation and paves the way for more cosmic experiments.\\ Utilizing Eq. (\ref{eq10}) and systematically substituting various values of the free parameters, we can identify the compatibility points between two conjectures through the numerical solution method. The presence of these compatibility points provides strong evidence for establishing the consistency between the two conjectures. However, beyond merely identifying these points, Eq. (\ref{eq10}) allows us to delineate the specific parameter ranges within which WGC and WCCC maintain compatibility. To facilitate a more structured analysis, we proceed by reformulating Eq. (\ref{eq10}) in the following manner,
\begin{equation}\label{eq11}
(\frac{Q^2}{M^2})_{ext}=1+\frac{2\gamma^{2/(3 \omega+1)}}{a^2(3\omega+1)(b-1)^{3\omega+1}}
\end{equation}
By classifying these ranges, we aim to establish a robust framework for evaluating the interplay between these two conjectures, offering deeper insights into their theoretical consistency and practical implications. In general, three distinct scenarios can be analyzed. So, with respect to Eq. (\ref{eq11}):\\\\
$1.\quad \omega>-\frac{1}{3},\quad a>0,\quad \gamma>0, \quad b>1$.\\\\ We encounter the establishing compatibility between the WGC and WCCC. To ensure that the parameter range encompasses values with positive confidence, we introduce an analytical expression in the form \( 3\omega+1=2n \). This formulation enables us to systematically exclude any negative probabilities, refining our analysis to focus exclusively on physically meaningful solutions. Given this refinement, we now proceed to elaborate on the distinct parameter intervals that dictate the compatibility conditions between WGC and WCCC.\\\\
$2. \quad \omega\neq\frac{2}{3}n-\frac{1}{3},\quad \omega<- \frac{1}{3},\quad a>0,\quad \gamma>0,\quad b<1$.\\\\ In this scenario, both expressions, finally, leads to a positive outcome, thereby establishing the consistency between the two conjectures.\\\\
$3. \quad \omega=\frac{2}{3}n-\frac{1}{3},\quad \omega>- \frac{1}{3},\quad a>0,\quad \gamma>0,\quad b<1$.\\\\ Within this range, both conjectures remain consistent with each other.
By choosing different black holes and classifying them, we can categorize black holes that can be investigated for the concept of quantum gravity. Extensive studies can then be started on these black holes, ensuring that other important concepts of physics are maintained, and the swampland program can be fully checked and tested.
\subsection{Discussion and result}
\begin{figure}[h!]
 \begin{center}
 \subfigure[]{
 \includegraphics[height=4cm,width=5.5cm]{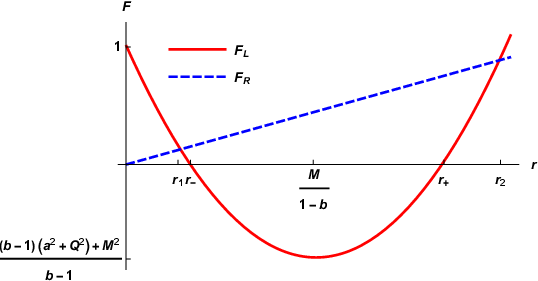}
 \label{3a}}
 \subfigure[]{
 \includegraphics[height=4cm,width=5.5cm]{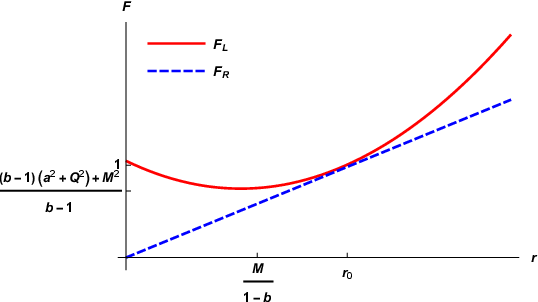}
 \label{3b}}
 \subfigure[]{
 \includegraphics[height=4cm,width=5.5cm]{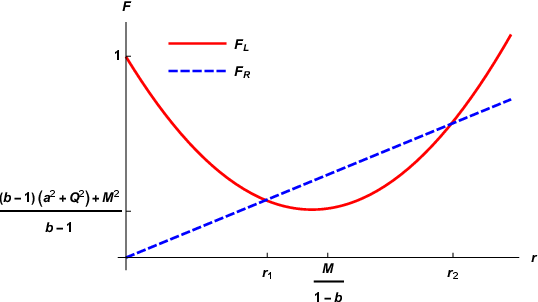}
 \label{3c}}
  \caption{\small{Fig. (\ref{m3}) shows the plot of $F-r$. $F_L=a^2+(1-b) r^2-2 m r+Q^2$ and $F_R =\gamma r^{1-3 \omega} $ with respect to $\omega=0$ and $a = 0.5; b = 0.1; \gamma = 0.5; \omega = 0$ for $M>Q$, $M=Q$, $M<Q$, respectively}}
 \label{m3}
 \end{center}
 \end{figure}

\begin{figure}[h!]
 \begin{center}
 \subfigure[]{
 \includegraphics[height=4cm,width=5.5cm]{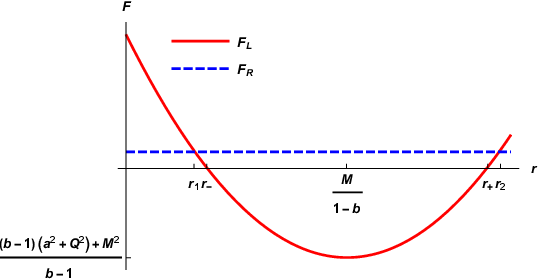}
 \label{4a}}
 \subfigure[]{
 \includegraphics[height=4cm,width=5.5cm]{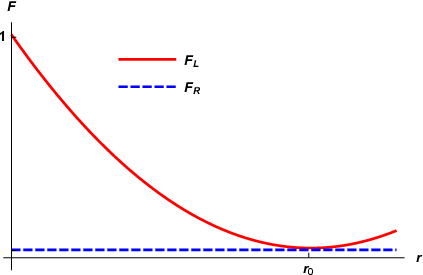}
 \label{4b}}
 \subfigure[]{
 \includegraphics[height=4cm,width=5.5cm]{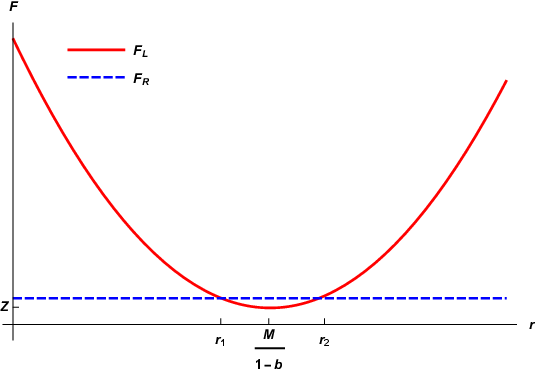}
 \label{4c}}
  \caption{\small{Fig. (\ref{m4}) shows the plot of $F-r$. $F_L=a^2+(1-b) r^2-2 m r+Q^2$ and $F_R =\gamma r^{1-3 \omega} $ with respect to $\omega=0$ and $a = 0.7; b = 0.1; \gamma = 0.2; \omega = 1/3$ for $M>Q$, $M=Q$, $M<Q$, respectively}}
 \label{m4}
 \end{center}
 \end{figure}

\begin{figure}[h!]
 \begin{center}
 \subfigure[]{
 \includegraphics[height=4cm,width=5.5cm]{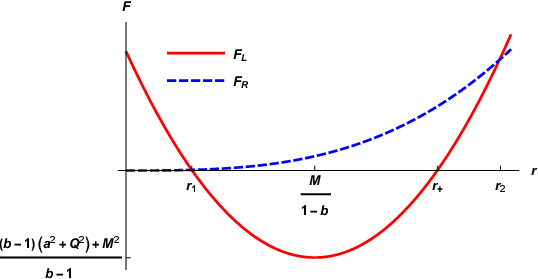}
 \label{5a}}
 \subfigure[]{
 \includegraphics[height=4cm,width=5.5cm]{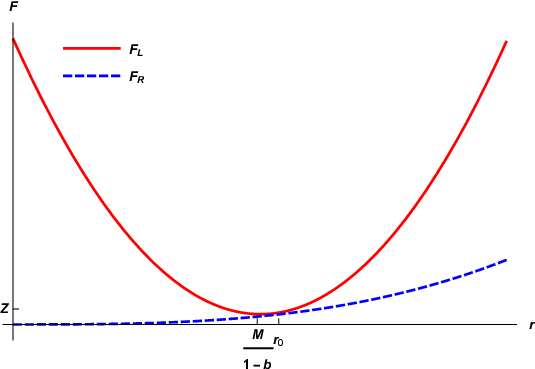}
 \label{5b}}
 \subfigure[]{
 \includegraphics[height=4cm,width=5.5cm]{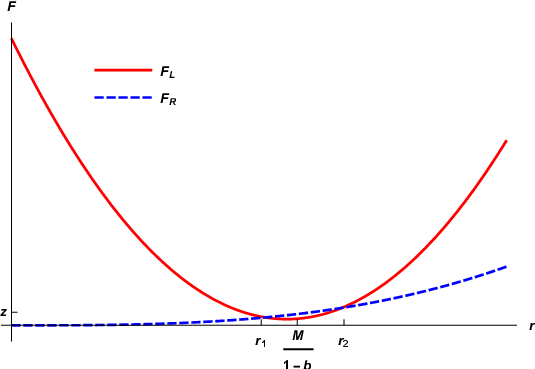}
 \label{5c}}
  \caption{\small{Fig. (\ref{m5}) shows the plot of $F-r$. $F_L=a^2+(1-b) r^2-2 m r+Q^2$ and $F_R =\gamma r^{1-3 \omega} $ with respect to $\omega=0$ and $a = 0.1; b = 0.15; \gamma = 0.1; \omega = -2/3$ for $M>Q$, $M=Q$, $M<Q$, respectively}}
 \label{m5}
 \end{center}
 \end{figure}

\begin{figure}[h!]
 \begin{center}
 \subfigure[]{
 \includegraphics[height=4cm,width=5.5cm]{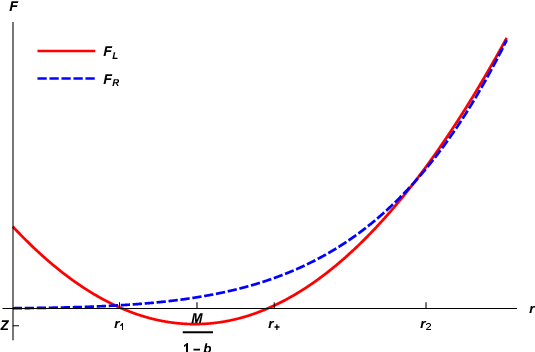}
 \label{6a}}
 \subfigure[]{
 \includegraphics[height=4cm,width=5.5cm]{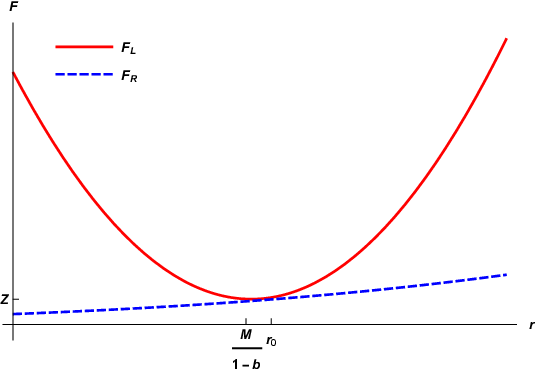}
 \label{6b}}
 \subfigure[]{
 \includegraphics[height=4cm,width=5.5cm]{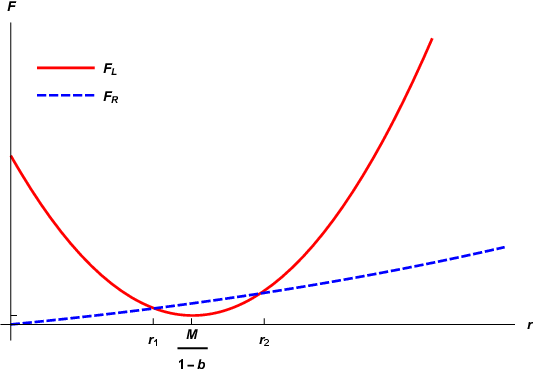}
 \label{6c}}
  \caption{\small{Fig. (\ref{m6}) shows the plot of $F-r$. $F_L=a^2+(1-b) r^2-2 m r+Q^2$ and $F_R =\gamma r^{1-3 \omega} $ with respect to $\omega=0$ and $a = 0.05; b = 0.05; \gamma = 0.2; \omega = -4/3$ for $M>Q$, $M=Q$, $M<Q$, respectively}}
 \label{m6}
 \end{center}
 \end{figure}

We investigate the behavior of the function \( F \) as a function of \( r \) for various values of \( M \), \( Q \), \( a \), \( b \), \( \gamma \), and \( \omega \). Refer to Fig. (\ref{m1}) for plots of \( F \) with different parameter choices. This figure illustrates how the metric function varies with \( a \), \( b \), \( \gamma \), and \( \omega \). The effects of these parameters are clearly visible in Fig. (\ref{m1}).
Analyzing the curve of \( F_L \), we find its minimum value is $((b-1) \left(a^2+Q^2\right)+M^2)/(b-1)$ at \( r_{min} = \frac{M}{1-b} \). In Figs. (\ref{3a}), (\ref{4a}), (\ref{5a}), and (\ref{6a}), we plot \( F_R \) and \( F_L \) for \( Q^2/M^2 \leq 1 \). The curves intersect at two points, \( r_1 \) and \( r_2 \), indicating that the KNKL black hole has an outer event horizon \( r_2 \) larger than the KNL black hole's outer event horizon \( r_+ \), and an inner event horizon \( r_1 \) smaller than the KNL black hole's inner event horizon \( r_- \). Conversely, in Figs. (\ref{3c}), (\ref{4c}), (\ref{5c}), and (\ref{6c}), we plot \( F_R \) and \( F_L \) for \( Q^2/M^2 > 1 \). Here, \( F_L \) does not cross the \( r \)-axis, meaning the ordinary KNL black hole has no event horizon and exposes its singularity, known as a naked singularity, violating the WCCC and invalidating the WGC. However, for the KNKL black hole, \( F_R \) and \( F_L \) still intersect at \( r_1 \) and \( r_2 \), even when \( Q^2/M^2 > 1 \), indicating two event horizons that conceal its singularity, thus not violating the WCCC and WGC.
In Figs. (\ref{3b}), (\ref{4b}), (\ref{5b}), and (\ref{6b}), \( F_R \) and \( F_L \) touch at a single point, defining the extremal state of the black hole at \( r_0 \) where \( Q^2/M^2 > 1 \), \( \lambda = \lambda_{ext} \), and \( b = b_{ext} \). Depending on the specific regions of these parameters, the black hole can either lead to a naked singularity, violating the WCCC, or satisfy both the WGC and WCCC simultaneously. This is illustrated in Figs. (\ref{m3}) to (\ref{m6}).\\ We plot \( F_L = a^2 + (1-b) r^2 - 2 M r + Q^2 \) and \( F_R = \gamma r^{1-3 \omega} \) in these figures, using Eq. (\ref{eq4}). Considering \( Q^2/M^2 \leq 1 \) and \( Q^2/M^2 > 1 \), in Figs. (\ref{3a}), (\ref{4a}), (\ref{5a}), and (\ref{6a}, when \( F_R = 0 \), \( F_L \) crosses the \( r \)-axis at \( r_+ \) and \( r_- \), the outer and inner event horizons of the standard KNL black hole.
We explore the conditions for WGC and WCCC compatibility for the KNKL black hole. The WGC and WCCC are distinct concepts in theoretical physics, addressing different aspects of gravity and spacetime. The WGC concerns the relative strengths of forces in quantum gravity, while the WCCC deals with singularities and event horizons in gravitational collapse within classical general relativity. Combining these ideas, one might speculate on potential connections between the WGC and cosmic censorship, suggesting that the WGC could influence cosmic censorship scenarios. This connection provides a bridge for better communication between general relativity and quantum mechanics, paving the way for further investigation and cosmic experiments.

\section{Conclusions}
WGC is a key component of the broader swampland program, which aims to differentiate between effective field theories (EFTs) that can be consistently integrated into a theory of quantum gravity (the landscape) and those that cannot (the swampland). The landscape is akin to a series of islands representing valid EFTs, while the swampland is a vast area of inconsistent EFTs. Within this framework, the WGC serves as a detector, providing criteria to identify theories that belong to the landscape. The WGC posits that in any effective field theory with a gauge symmetry (such as U(1)), there must exist a particle whose charge-to-mass ratio exceeds a certain threshold, typically expressed as ( $q/m > 1$ ). This ensures that gravity remains the weakest force at sufficiently high energy scales. There are some types of charged black holes: subextremal $(Q < M)$. Extremal (Q = M): Here, the inner (Cauchy) horizon and the outer (event) horizon merge into a single horizon. Superextremal $(Q>M)$: The rarest type, where the event horizon disappears entirely, leaving a naked singularity. Also, there are some types of particles: subextremal Particles: These particles have a charge-to-mass ratio of less than 1. Superextremal Particles: These particles have a charge-to-mass ratio greater than 1, meaning they are more "charged" than their mass would typically allow. The relationship between the WGC and black holes primarily revolves around the stability and decay processes of extremal black holes. If the WGC is satisfied, extremal black holes can decay by emitting superextremal particles, thus preventing the formation of naked singularities and preserving the cosmic censorship hypothesis. If the WGC is satisfied: Extremal black holes can decay by emitting superextremal particles, preventing the formation of naked singularities. If the WGC is violated: An extremal black hole could only shed its charge by emitting subextremal particles, potentially leading to a superextremal black hole and introducing a naked singularity into spacetime, violating cosmic censorship. Therefore, the WGC acts as a safeguard against such violations by ensuring that the decay processes of black holes are consistent with the principles of quantum gravity. One of the factors required for a black hole to undergo its natural collapse process until complete evaporation is the presence of superextremal particles. The likelihood of such particles existing in subextremal and extremal black holes is zero or significantly lower compared to superextremal black holes, which generally exhibit naked singularities.
Identifying models that exhibit the black hole in a superextremal form reinforces the validity and necessity of the WGC. A superextremal black hole, with a charge-to-mass ratio greater than one, necessitates the existence of superextremal particles. If such a black hole can exist without creating observable naked singularities, it indicates a stable balance of forces that constitute black hole mechanics\cite{321a}.
To analyze a black hole with general behavior, various conditions such as the presence of an event horizon, the WCCC, stability, thermodynamic properties, and more can be examined.\\\\
According to the above explanations, to find more evidence for the WGC, we must study various cosmic phenomena and contradictions, and compare them with existing observable data. This approach leads to a special classification of black holes from the perspective of the WGC. Another important aspect is understanding and applying thermodynamic relationships and respecting the laws of thermodynamics. This work creates a bridge between quantum mechanics and cosmology through the investigation of the WGC and the concept of cosmic censorship.
To gather more evidence for the WGC, researchers need to examine a wide range of cosmic cases. This involves identifying and analyzing contradictions within observable data. By doing so, we can classify black holes in a new way, specifically from the viewpoint of the WGC. This classification helps in understanding how different types of black holes behave under the conjecture's framework.
This work addresses the potential validation of the WGC alongside the WCCC, presenting a significant challenge in quantum gravity. By exploring the KNKL black hole, we establish a surprising connection between these seemingly unrelated conjectures, thereby bridging the quantum and cosmic realms. Our findings demonstrate that by imposing specific constraints on the black hole's parameters, the WGC and WCCC can be compatible in certain regions. We examined the properties of the KNKL black hole for \(Q > M\) and presented figures to test the validity of both conjectures simultaneously. Without the spin parameter \(a\), cloud of string parameter \(b\), quintessence parameter \(\gamma\), and equation of state parameter \(\omega\), the black hole either has two event horizons if \(Q^2/M^2 \leq 1\) or none if \(Q^2/M^2 > 1\), leading to a naked singularity that contradicts the WCCC. However, when \(a\), \(b\), \(\gamma\), and \(\omega\) are present, the black hole has event horizons in some regions for \(Q^2/M^2 > 1\), ensuring the singularity is covered and both the WGC and WCCC are fulfilled. In the extremality state of the black hole, these conjectures remain viable, with the black hole maintaining an event horizon. We conclude that certain regions of \(a\), \(b\), \(\gamma\), and \(\omega\) parameters can make the WGC and WCCC compatible, indicating their agreement when these parameters are present. If there is a black hole that has an event horizon at the extreme level and the WGC is also valid, we have consistency between WGC with WCCC. Conversely, if at a point where the WGC is valid for the extremal black hole, it seems that the WGC also takes on the role of WCCC, and the structure has an event horizon.\\\\ Also, we face some questions that we leave to future works:\\\\1. Can the compatibility of the WGC and WCCC be extended to higher-dimensional black holes, and if so, what additional constraints or modifications are necessary?\\
2. How do quantum effects influence the compatibility of the WGC and WCCC in the context of the KNKL black hole?\\
3.  Are there observable astrophysical phenomena that could provide empirical evidence supporting the compatibility of the WGC and WCCC with the parameters \(a\), \(b\), \(\gamma\), and \(\omega\)?\\
4.  How do alternative theories of gravity impact the relationship between the WGC and WCCC in black hole solutions?\\\\
The implications of finding compatibility between the WGC and the WCCC in the context of the KNKL black hole are profound and multifaceted.
 Demonstrating that the WGC and WCCC can coexist under certain conditions suggests a deeper, unified framework in which quantum gravity and general relativity principles are harmonized. This could pave the way for new theories that integrate these two fundamental aspects of physics.
Understanding the conditions under which these conjectures are compatible enhances our knowledge of black hole physics, particularly in complex scenarios involving additional parameters like spin, string clouds, and quintessence. This could lead to more accurate models of black hole behavior and evolution.
The connection between the WGC and WCCC may provide insights into the nature of quantum gravity, potentially offering clues about how gravity operates at the quantum level and how it interacts with other fundamental forces.
The results could inspire new theoretical developments and explorations in higher-dimensional black holes, alternative theories of gravity, and the role of quantum effects in black hole dynamics.
Overall, these findings open up exciting avenues for research, potentially leading to breakthroughs in our understanding of the universe's most extreme environments and the fundamental nature of gravity.

\section{Acknowledgments}
The work of Saeed Noori Gashti is supported by the Iran National Science Foundation (INSF). This work is based upon research funded by Iran National Science Foundation (INSF) under project No.4038260. The authors would like to express their gratitude to Mohammad Reza Alipour for his insightful discussions.

\end{document}